\documentstyle[aps,prl,epsfig,multicol]{revtex}

\begin{document}
\title{Magnetization plateau and quantum phase transitions in a
spin-orbital model}
\author{Zu-Jian Ying$^{1,2,3}$, Angela Foerster$^2$, Xi-Wen Guan$^4$, 
Bin Chen$^1$, Itzhak Roditi$^3$}
\address{1. Hangzhou Teachers College, Hangzhou 310012, China\\
2. Instituto de F\'\i sica da UFRGS, Av. Bento Gon\c calves, 9500, Porto
Alegre, 91501-970, Brasil\\
3. Centro Brasileiro de Pesquisas F\'\i sicas, Rua Dr. Xavier Sigaud 150,
22290-180 Rio de Janeiro, RJ, Brasil\\
4. Department of Theoretical Physics, 
Research School of Physical Sciences and Engineering,
and Centre for Mathematics \\ and its Applications,
Mathematical Sciences Institute,
Australian National University, Canberra ACT 0200,  Australia
}
\date{\today }
\maketitle

\begin{abstract}
A spin-orbital chain with different Land\'e $g$ factors and one-ion
anisotropy is studied in the context of the thermodynamical Bethe
ansatz.
It is found that there exists a
magnetization plateau resulting from the different Land\'e $g$ factors. 
Detailed phase
diagram in the presence of an external magnetic field is presented 
both numerically
and analytically. For some values of the anisotropy, the
four-component system undergoes five consecutive quantum phase transitions
when the magnetic field varies.  We
also study the magnetization in various cases, especially its behaviors
in the vicinity of the critical points.
For the SU(4) spin-orbital
model,  explicit analytical expressions for the critical fields are
derived, with excellent accuracy compared
with numerics.
\end{abstract}


\begin{multicols}{2}

Orbital degeneracy in electron systems leads to rich and novel magnetic
phenomena in many transitional metal oxides\cite{Nagaosa}. Among them
are the orbital
ordering and orbital density wave, 
which have been observed experimentally in a
family of manganites\cite{Saitoh}. A tractable model to describe 2-fold
orbital degenerate system is the SU(4) model\cite{LiSU4}, which
has attracted much attention\cite{LiSU4,LiSU4BA,Yamashita,SJGu,Mila,%
Azaria,Frischmuth}. In 
the one-dimensional case the model is exactly solvable 
by Bethe ansatz (BA)\cite{LiSU4BA}. 
An interesting question is to study the critical behavior of such a
system in an external magnetic field, especially when
different Land\'e $g$ factors for spin
and orbital sectors are involved. One may expect that the difference of $g$ 
factors will bring about new physics
as a result of the competition of the spin and orbital degrees of
freedom. 
In Ref.\cite{Yamashita}, the authors studied the magnetic properties of
the SU(4) model via BA, without taking different $g$ factors into account,
whereas
numerical calculation was performed 
in Ref.\cite{SJGu} for the model with different $g$ factors
for up to 200 lattice sites.
However, a full picture
about the critical fields is still lacking.
Another motivation is to see whether or not
any magnetization plateau (MP), an
interesting magnetic phenomenon, occurs
in such a spin-orbital model. 
As is well known, antiferromagnetic chains
with integer spin are gapful\cite{Haldane}, whereas for half-integer spin there
also exists a gapful phase with a MP in the presence of a large planar
anisotropy\cite{Oshikama}. Also fractional MP have been observed and can be
explained by Shastry-Sutherland lattice\cite{FracMz}. But an MP arising from
different Land\'e g factors has not been addressed yet.

Deviation from the SU(4) symmetry was considered by variation in the interaction
parameters of neighbor sites\cite{Mila,Azaria}, while another possible deviation
may result from the one-ion interaction. Since many compounds are
magnetically anisotropic in which the orbital angular momentum (OAM) may be
constrained in some direction due to crystalline field, the angle between
spin and OAM determines the spin-orbital coupling (SOC) energy. This kind of
one-ion SOC leads to magnetic anisotropy\cite{Slonczewski}%
. Under the influence of molecular field and an external
field, the spin is parallel to the OAM, then the SOC energy depends on
whether they are in the same directions. In such a case, $%
s_i^z\tau _i^z$ type of interaction describes well the SOC energy.
Here we introduce
such an SU(2)$\otimes $SU(2) SOC interaction into the SU(4) model.
A detailed investigation of the phase diagram is undertaken both numerically 
and analytically in the context of the thermodynamical Bethe ansatz (TBA).
We find that the system exhibits an
MP resulting from different $g$ factors when the SOC is sufficiently strong.
The critical behavior of the magnetization in the vicinities of the
critical points is revealed.
For certain values of Land\'e $g$ factor, the model undergoes five consecutive 
quantum phase transitions when the external magnetic
field varies. Further, the explicit analytic expressions for the
critical fields for the SU(4)
model are derived, with excellent accuracy compared to numerical results.

{\it The model and TBA.} Consider an L-site chain with the Hamiltonian 
\begin{eqnarray}
{\cal H} &=&{\cal H}_0+{\cal H}_z+{\cal M},\quad {\cal H}_0=%
\sum_{i=1}P_{i,i+1},  \nonumber  \\
{\cal H}_z &=&\Delta _z\sum_is_i^z\tau _i^z,\ {\cal M}=-g_sH\sum_is_i^z-g_tH%
\sum_i\tau _i^z,  \label{H}
\end{eqnarray}
where $\vec s$ and $\vec \tau $ are spin-1/2 operators for spin and orbital
sectors, and $g_s$ and $g_t$ denote, respectively, the Land\'e $g$ factors. 
We assume
$g_s>g_t$ throughout the paper.  ${\cal H}_0$ is the SU(4) model with $%
P_{i,j}=\left( 2\vec s_i\cdot \vec s_j+ 1/2 \right) \left( 2\vec \tau
_i\cdot \vec \tau _j+ 1/2 \right) $ exchanging the four site states $\left|
s_i^z\tau _i^z\right\rangle $: $\phi _1=\left| \uparrow \downarrow
\right\rangle $, $\phi _2=\left| \downarrow \uparrow \right\rangle ,$ $\phi
_3=\left| \uparrow \uparrow \right\rangle $, $\phi _4=\left| \downarrow
\downarrow \right\rangle $. The symmetry is broken into SU(2)$\otimes $SU(2)
by ${\cal H}_z$ and further into four U(1)'s 
by the external magnetic field $H$. The model can be
solved exactly via BA approach. The BA equations are the same as the SU(4)
model\cite{LiSU4BA,Sutherland} under the periodic boundary conditions,
with the energy eigenvalues given by
\begin{eqnarray}
E=-2 \pi \sum_{i=1}^{M^{(1)}} a_1(\lambda _i)+\sum_{k=1}^4E_kN_k,
\end{eqnarray}
where $a_n\left( \lambda \right) = 1/ (2\pi) \; n / (\lambda ^2+n^2/4)$, 
and $E_1=-\Delta _z/4-g_{-}H/2,\ E_2=-\Delta _z/4+g_{-}H/2$, $E_3=\Delta
_z/4-g_{+}H/2,\ E_4=\Delta _z/4+g_{+}H/2$, with $g_{\pm }=g_s\pm
g_t$. $N_k$ is the total site number in state $\phi _k$ and $M^{(i)}$ ($%
i=1,2,3$) is the rapidity number. For a certain choice of the basis order,
which depends on whether or not the component is energetically
favorable, the energy
can be rewritten as $E=\sum_{i=1}^{M^{(1)}}g^{(1)}(\lambda
_i)+g^{(2)}M^{(2)}+g^{(3)}M^{(3)}$, with the coefficients $g^{(i)}$
depending on the basis order. In the thermodynamical limit, 
the summation over the rapidity
becomes integral. Following Refs.\cite
{string} and \cite{Y-Y}, one may obtain the ground state (GS)
equations for the dressed energies $\epsilon ^{(i)}\;(i=1,2,3)$, 
\begin{equation}
\epsilon ^{(i)}=g^{(i)}-a_2*\epsilon ^{(i)-}+a_1*(\epsilon
^{(i-1)-}+\epsilon ^{(i+1)-}),  \label{TBA}
\end{equation}
where $\epsilon ^{(0)}=\epsilon ^{(4)}=0$
and the symbol $*$ denotes the convolution. The dressed energies
describe elementary excitations over the Fermi seas, i.e., the GS with
all states with negative dressed energies filled.
According to an energetics argument, we may divide the external field $H$ into
three regions: (I) $0\leq H<H_{R1}$, (II) $H_{R1}<H<H_{R2}$, (III) $%
H_{R2}<H<\infty $ with $H_{R1}=\left| \Delta _z\right| /(2g_s),\quad
H_{R2}=\left| \Delta _z\right| /(2g_t)$. For $\Delta _z>0,$ the
corresponding basis order are: (I$_{+}$) ($\phi _1$,$\phi _2,\phi _3,\phi _4$%
)$^T$, (II$_{+}$) ($\phi _1$,$\phi _3,\phi _2,\phi _4$)$^T$, (III) ($\phi _3$%
,$\phi _1,\phi _2,\phi _4$)$^T$; for $\Delta _z<0$: (I$_{-}$) ($\phi _3,\phi
_4,\phi _1$,$\phi _2$ )$^T$, (II$_{-}$) ($\phi _3,\phi _1$,$\phi _4,\phi _2$)%
$^T$, (III) the same as $\Delta _z>0$. These five basis orders provide a
full description of the
phase diagram of the system.

\begin{figure}[t]
\setlength\epsfxsize{70mm}
\epsfbox{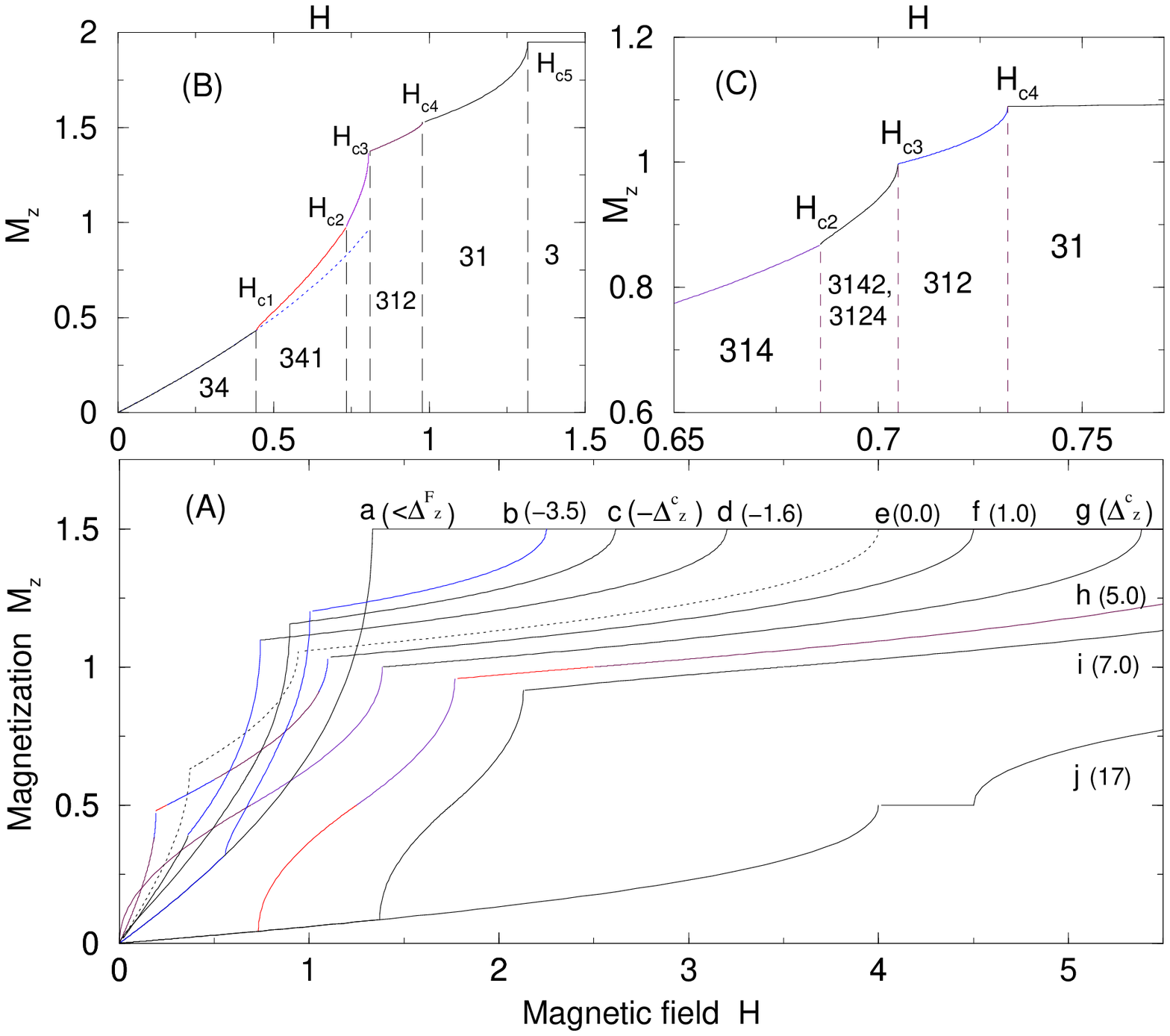}
\caption{(A) Typical magnetization behaviors for fixed values of Land\'e $g$
factors $g_s=2.0$ and $g_t=1.0$. The numbers in
brackets indicates values of the anisotropy parameter $\Delta _z$. 
The dotted line e denotes the magnetization for the SU(4) model.
Curve j exhibits a magnitization plateau. 
(B) Magnetization corresponding to five consecutive five $H_c$
quantum phase transitions for $g_s=2.0$, $g_t=1.9$, and $\Delta _z=-3.0$. 
Here the variation
of the state component numbers is 
2$\rightarrow $3 $\rightarrow $4$\rightarrow $%
3$\rightarrow $2$\rightarrow $1. 
The number $i$
labels the state $\phi _i$, e.g., the state components in 
the phase 123 are $\phi
_1\phi _2\phi _3$ in which $\phi _1$ is the most energetically favorable
whereas $\phi _3$ is the least energetically favorable.
The phase variations between $H_{c2}$ and $H_{c3}$ are 3412$\rightarrow $3142%
$\rightarrow $3124. The dotted line for comparison is
an extension of the phase 34 by assuming the components unchanged. 
(C) Magnetization for 
consecuitive five $H_c$ phase transitions for $g_s=2.0$ and $g_t=1.0$, 
with $\Delta _z=-1.4$. $H_{c1}=$%
0.531 (transition 3142$\rightarrow $214) and $H_{c5}=3.3$3 (31$\rightarrow $3)
are relatively far away. The variation of the component numbers in the 
phase transitions is 4%
$\rightarrow $3$\rightarrow $4$\rightarrow $3$\rightarrow $2$\rightarrow $1.
}
\label{Mz}
\end{figure}

{\it Plateau.} The competition between the anisotropy parameter $%
\Delta _z$ and the magnetic field $H$ results in a novel quantum phase
diagram. In the absence of the magnetic field, it is easy to find that the
states $\phi _3$ and $\phi _4$ are gapful for $\Delta _z>\Delta _z^c=4\ln 2$%
. Whereas for $\Delta _z<-\Delta _z^c$, the components $\phi _1$ and $\phi
_2 $ are gapful. Therefore, the GS is in an $su(2)$ spin-orbital liquid
state in strong anisotropy regime in the absence of the field. However, 
the presence of the magnetic field completely splits 
all four components energetically.
The magnetization $M^z=g_ss^z+g_t\tau ^z$ increases from
zero. For large positive $\Delta _z$, the
field bring the component $\phi _3$ closer to the GS, while the component $%
\phi _2$ gradually gets out of the GS. Certainly, if the field reaches the
first critical field where the component $\phi _3$ has not yet involved in
the GS, a quantum phase transition from the spin-orbital liquid phase to a
ferromagnetic phase occurs. Thus a magnetization plateau opens with a
constant magnetization $M^z=g_{-}/2$. Nevertheless, this plateau will
disappear when the field is strong enough $H>H_{c2}^p$. In such a case,
the component $\phi _3$
becomes involved in the GS. The critical field $H_{c2}^p$ indicates a
quantum phase transition from the ferromagnetic GS into a spin-orbital
liquid phase. If the field continues to increase, the spin and orbital
sectors
become fully-polarized at the third critical point $H_{c3}^p$. From the TBA
equations (\ref{TBA}), we get the exact expressions for the critical
fields 
\begin{equation}
H_{c1}^p=\frac 4{g_{-}},\quad H_{c2}^p=\frac{\Delta _z/2-4}{g_t},\quad
H_{c3}^p=\frac{\Delta _z/2+4}{g_t}.  \label{Hc}
\end{equation}
Notice that the plateau opens only if $\Delta _z>\Delta _z^P=8g_s/g_{-}\
$ and $0<g_t<g_s$. If the $g$ factors are the same, the plateau disappears
because the components $\phi _1$ and $\phi _2$ remain degenerate in 
the field. The
critical behavior of the magnetization in the vicinities 
of the critical points may be summarized as follows
\begin{equation}
\left\langle M^z\right\rangle \cong \left\langle M^z\right\rangle _c+\eta
k_M\delta H^{\frac 12},  \label{Mz_p}
\end{equation}
where $\left\langle M^z\right\rangle _c=g_{\mp }/2$ are, respectively,
plateau and saturation magnetizations. $\delta H$ is the small deviation
from the critical points and $\eta =\pm 1$ depending on $M^z$ is increasing or
decreasing. $k_M=g_{-}^{3/2}/\pi $ near $H_{c1}^p$ and $%
k_M=g_t^{3/2}/\pi $ near $H_{c2}^p$ and $H_{c3}^p$. The coefficients $g_{-}$
and $g_t$ in (\ref{Hc}) and (\ref{Mz_p}) can be easily understood, since only 
$\phi _1$ and $\phi _2$ exist in the GS before $H_{c1}^p$ is reached, 
the differences of
their field energy and magnetization are $g_{-}H$ and $g_{-}$ respectively.
While for $H_{c2}^p<H<H_{c3}^p$, only $\phi _1$ and $\phi _3$ compete in the
GS, since $\phi _2$ already gets out
before $\phi _3$ enters the GS. $\phi _1$ and $\phi _3$ differ in field
energy by $g_tH$ and in magnetization by $g_t$. 
In Fig.\ref{Mz}A we plot
the magnetization curves for different values of the parameter $\Delta _z$, 
including a
plateau case.

\begin{figure}
\setlength\epsfxsize{70mm}
\epsfbox{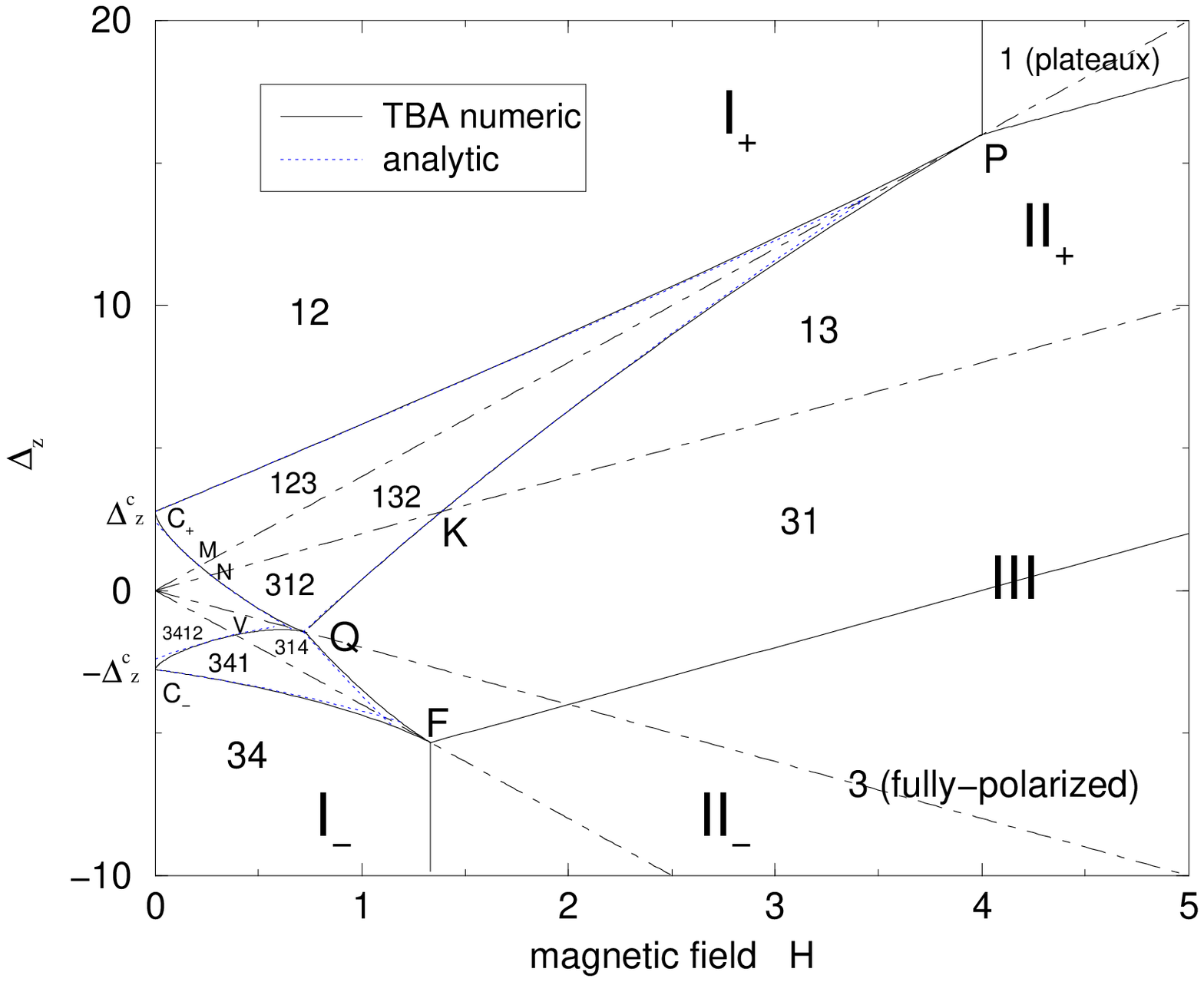}
\caption{Phase diagram for fixed values of Land\'e $g$ factors $g_s=2.0$ 
and $g_t=1.0$. 
Inside C$_{+}$QC$_{-}$,
the 4-state phases are 1234 (I$_{+}$), 1324 (II$_{+}$), 3124 (III), 3142 (II$%
_{-}$), and 3412 (I$_{-}$). The discrepancy of the analytic curves from the
numerical ones
is not visible for most regions of C$_{+}$PQ and C$_{+}$MNQ (with typical
differences within 1.0\%, and 0.1\%, respectively). There is less accuracy in
regions I$_{-}$ and II$_{-}$ for the analytic results due to larger driving
terms and smaller Fermi boundaries. Magnetization plateaux 
and fully-polarized cases are exact.}
\label{phase}
\end{figure}

{\it The phase diagram: numerical and analytical.} Here we present a
detailed analysis of the GS phase diagram
both numerically and analytically.
In Fig.\ref{phase}, we plot the phase diagram with respect to the
parameters $\Delta _z$ and $H$ for fixed values of $g_s$ and $g_t$
($g_s =2.0$ and $g_t=1.0$). For convenience, we refer to the GS with 
$i$ components as $i$-state GS. 
Then for the phase transition between 3-state and
2-state GS, the critical fields follow from the Wiener-Hopf method\cite{WH},
which is valid for large Fermi boundaries. Explicitly, we have
\begin{eqnarray}
&&H_c^{PC_{+}}\doteq (\Delta _z-\Delta _z^c)g_{+}^{-1}-\tau
_1g_{-}^2g_{+}^{-3}(\Delta _z-\Delta _z^c)^2,  \nonumber   \\
&&H_c^{QP}\doteq \frac{\Delta _z^c+\frac{\Delta _z}2}{2g_s-g_t}+\tau _1%
\frac{g_{-}^2(\Delta _z-\Delta _z^K)^2}{(2g_s-g_t)^3},  \nonumber \\
&&H_c^{QF}\doteq \frac{\Delta _z^c-\frac{\Delta _z}2}{2g_s+g_t}+\tau _1%
\frac{[g_t\Delta _z^c-(g_s+g_t)\Delta _z]^2}{(2g_s+g_t)^3},  \nonumber \\
&&H_c^{FC_{-}}\doteq \pi ^2{g}_{-}{g}_{+}^{{-2}}[\sqrt{{1-\tau _14{g}%
_{-}^{-2}{g}_{+}^2(\Delta _z+\Delta _z^c)}}{-1]}, \label{H1}
\end{eqnarray}
where $\tau _1=1/(2\pi ^2)$ and $\Delta _z^K=2g_tH^K=g_t/g_{-}\Delta
_z^c $ which corresponds to infinite Fermi boundaries. Point Q is determined
by $\Delta _z^Q=-2g_tH^Q$,
with $H^Q\doteq $ $2\ln 2/g_s+4\pi ^{-2}\ln ^22g_t{}^2/g_s^3$. Near P or F the
above analytic results deviate due to small Fermi boundaries. 
But in this case, an analysis may be carried out in terms of expansion
of small Fermi boundaries. This leads us to
$H_c^{PK},H_c^{PC_{+}}\cong \frac{\Delta _z}{2g_s}\pm (g_{-}/2)^{%
\frac 32}/(\pi g_s^{\frac 52})(\Delta _z^P-\Delta _z)^{\frac 32}$. 
Similarly, near F, 
we have $H_c^{FQ}$, 
$H_c^{FC_{-}}\cong -\frac{\Delta _z}{2g_s}\pm (g_{+}/2)^{\frac 32%
}/(\pi g_s^{\frac 52})(\Delta _z-\Delta _z^F)^{\frac 32}$.

For $\left| \Delta _z\right| <\Delta _z^c$, the GS involves all the
four components, the magnetic field first brings about a phase transition 
from a 4-component liquid to a 3-component liquid at the phase boundary
C$_{+}$QC$_{-}$. Here one of the four components, which is energetically
unfavorable, completely gets out of the
GS, then the corresponding Fermi sea disappears. 
So the critical field only involves two
Fermi points $B_1$ and $B_2$. At point C$_{+}$, the component $\phi _2$ is
degenerate with the component $\phi _1$ which is the most energetically
favorable, the Fermi point $B_1$
lies at infinity.
Increasing $H$ along C$_{+}$MNQ drives $\phi _2$ away from $\phi _1$%
, so $\phi _2$ becomes less energetically favorable in the GS. 
Therefore the first Fermi sea
shrinks, i.e., $B_1$ decreases from infinity. Beyond M point, both the
increase of $H$ and the decrease of $\Delta _z$ make $\phi _3$ sink below the
$\phi _2$ which is rising, the energy difference between $\phi _3$ and $\phi _1$
begins to dominate over $B_1$. As $\phi _3$ is drawing near $%
\phi _1$, the first Fermi sea becomes broadened again with an increase
of $B_1$. After point N, $\phi _3$ sinks beyond $\phi _1$ to be the lowest
state, $\phi _1$ becomes less favorable in the GS. The first Fermi sea
shrinks again, $B_1$ begins to decrease along NQ from the infinity at N.
A similar analysis is applicable to $B_2$, which is mainly influenced by the
energy difference between the second and third components in energy levels. $%
B_2$ rises from zero at C$_{+}$ to infinity at M and decreases along MNQ to
zero at Q.  In the respective sections 
of C$_{+}$MNQ the critical fields take the form
\begin{eqnarray*}
&&H_c^{C_{+}MNQ}\doteq \Delta
_{a,1}/g_{+}+H_{B_2}^{C_{+}MNQ}+wH_{B_1}^{C_{+}MNQ} \\
&& \stackrel{MC_{+}}{=}H_\infty ^{+}+\tau _2\Delta _{a,3}^2/g_{+}+w\tau
_2(g_{-}a_0-g_t\Delta _{a,3})^2/g_{+}^3 \\
&& \stackrel{MN}{=}H_\infty ^{+}+\tau _2[(g_{-}a_0+g_s\Delta
_{a,3})^2+w(g_{-}a_0-g_t\Delta _{a,3})^2]/g_{+}^3 \\
&& \stackrel{NQ}{=}H_\infty ^{+}+\tau _2(g_{-}a_0+g_s\Delta
_{a,3})^2/g_{+}^3+w\tau _2\Delta _{a,3}^2/g_{+},
\end{eqnarray*}
where $\Delta _{a,m}=a_0-\frac{m\Delta _z}2$, $a_0=\frac{\sqrt{3}}2\pi -%
\frac 32\ln 3$, $\tau _2=\frac 3{16\pi ^2}$, and $w=2/3$. In each case,
the first
term comes from infinite $B_1$ and $B_2$, the second term is correction
from finite but large $B_2$, and the third term is the leading 
correction from the
larger $B_1$. For C$_{+}$MN near point M the $B_2$ and $B_1$ terms need to
be exchanged since $B_2$ becomes larger. Along NQ $B_1$ is always 
larger than $B_2$. The
location of $B_1=B_2$ in C$_{+}$MN may be estimated by $H_{B_1}=H_{B_2}$,
which gives $\Delta _z=2a_0(2g_s-g_t)/(3g_s)$ for C$_{+}$M and $\Delta _z=%
\frac 23a_0$ for MN. This coincides well with numerics, e.g., for MC$_{+}$
and $g_s=$2.0$,$ $g_t=$1.0, the analytic result is $\Delta _z=$1.073 whereas
the numerical one is 1.042.

Similarly, for C$_{-}$VQ, the term resulting from the infinite Fermi boundaries
is $H_\infty ^{-}=\Delta
_{a,-1}/g_{-}$, the correction terms are $H_{B_2}^{VC_{-}}=\tau _2\Delta
_{a,-3}^2/g_{-}$, $H_{B_1}^{VC_{-}}=H_{B_1}^{VQ}=\tau _2[g_{+}a_0+g_t\Delta
_{a,-3}]^2/g_{-}^3$, and  $H_{B_2}^{VQ}=\tau _2[g_{+}a_0+g_s\Delta
_{a,-3}]^2/g_{-}^3$, respectively. These expressions are not valid 
for $g_s\sim g_t$ due to small Fermi boundaries.

When the system is fully-polarized, only $\phi _3$ exists in the GS, while
the other components are all gapful, with a gap $\Delta =\min
\{E_i-4-E_3\mid i=1,2,4\}$. This gap is closed if $H<H_f$, with the
fully-polarized critical point 
\begin{equation}
H_f=\max \left\{ \left( \Delta _z/2+4\right) g_t^{-1},\ 4 g_{+}^{-1}\right\}. 
\label{Hfull}
\end{equation}
This expression is exact and valid for all $\Delta _z$. When $\Delta _z\leq
\Delta _z^F=-8g_s/g_{+}$, the strong negative anisotropy makes $\phi _1$ and 
$\phi _2$ too far away from $\phi _4$. Before the field
brings them close enough to get involved in the GS, the component $\phi _4$
has been all pumped out by the field from the GS at critical point $%
H_f=4/g_{+}$. For all $\Delta _z\leq \Delta _z^F$, the magnetization is
the same as shown by curve a in Fig.\ref{Mz}A.

The analytic results are compared with the numerics in Fig.\ref{phase},
with very satisfactory accuracy.

\begin{figure}
\setlength\epsfxsize{70mm}
\epsfbox{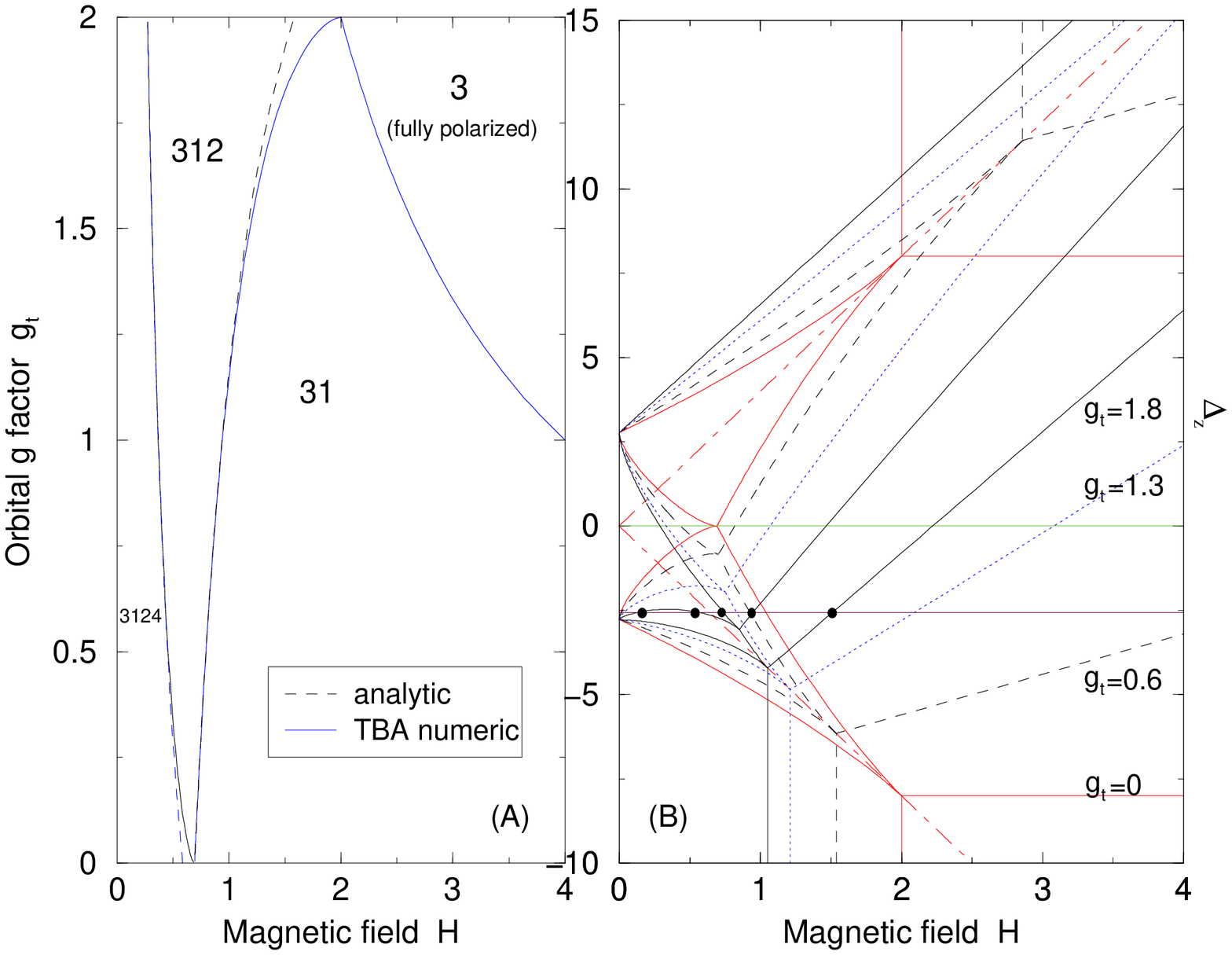}
\caption{(A) Comparison of the analytic and the numerical results for the
three critical fields of the SU(4) model with respect to the orbital $g$
factor $g_t$ for a fixed value of the spin $g$ factor $g_s=2.0$.
The numbers label the state components as in Fig.\ref{phase}, and $H_{c3}$
(phase transition 31$\rightarrow $3) is exact. (B) Phase diagrams for
various $g_t$ ($g_s=2.0$). 
$\Delta _z=0$ corresponds to the SU(4) model. The black dots
mark consecutive five $H_c$ phase transitions. Such a phenonemon exists 
for all $0<g_t<g_s $,
with $H_{c1}$-$H_{c4}$ being closer for smaller $g_t$.
}
\label{SU4gt}
\end{figure}

{\it Five consecutive $H_c$ phase transitions.} The competition of 
anisotropy $\Delta _z$
and the external field $H$ also leads to an unusual magnetic phenomenon. For
some fixed values of $\Delta _z$, $g_s$ and $g_t$, the system undergoes five
consecutive quantum phase transitions when $H$ varies, 
although the 4-component model usually
has at most three consecutive phase transitions. 
A strong negative anisotropy $\Delta _z$
makes $\phi _4$ energetically quite favorable. 
The field $H$ expels $\phi _2$ first
from the 4-component GS. However, before $H$ overwhelms the influence of $%
\Delta _z$ on $\phi _4$, further increase of $H$ will make $\phi _2$ closer
to $\phi _4$ and draw it back into the GS. This process brings about the
first two phase transitions. Then $H$ plays a dominant role, it begins to
bring out $\phi _4$, $\phi _2$ and $\phi _1$ from the GS one by one.
This results in other three consecutive phase transitions. 
The variation of the state
component numbers in the GS is: 4$\rightarrow $3$\rightarrow $4$\rightarrow 
$3$\rightarrow $2$\rightarrow $1. This five $H_c$'s case exists for all $%
0<g_t<g_s $ and becomes more visible when $g_t$ is larger. 
One case is marked by black dots in Fig.\ref{SU4gt}B for $g_s=2.0$ 
and $g_t=1.8$. Another
possible case for five $H_c$ transitions to occur is
the GS composed of $\phi _3$ and $%
\phi _4$. The field brings $\phi _1$ into the GS first. As
the difference of $g_t$ and $g_s$ is getting smaller, $\phi _2$ has closer
energy to $\phi _1$.  Further increase of $H$  brings $\phi _2$ into
the GS before $\phi _4$ completely gets out. The component number changes in
such a way: 2$\rightarrow $3$\rightarrow $4$\rightarrow $3$\rightarrow $2$%
\rightarrow $1. This case occurs when the point Q is below C$_{\_}$ in Fig.%
\ref{phase}, approximately requiring $1>g_t/g_s>1-2\ln 2g_t^3/(\pi ^2g_s^3)$%
. Four consecutive $H_c$ transitions take place 
for $\Delta _z=-\Delta _z^c$. In Fig.\ref{Mz}%
B and C, we plot the magnetization curves which display five consecutive
$H_c$ phase transitions.

{\it The SU(4) model.} If we set the anisotropy parameter $\Delta _z$ to
be zero, the model reduces to the $SU(4)$ model with different Land\'e
$g$ factors in the spin and orbital sectors. In this special case, the
above results for
the critical fields give rise to 
\begin{eqnarray}
H_{c1}^{SU(4)} &\doteq &a_0 g_{+}^{-1}+\tau _2a_0^2(2g_s-g_t)^2g_{+}^{-3}
+w\tau_2 a_0^2 g_{+}^{-1},  \nonumber \\
H_{c2}^{SU(4)} &\doteq &\Delta _z^c\ (2g_s-g_t)^{-1}+\tau _1
(g_t\Delta _z^c)^2(2g_s-g_t)^{-3}, \\
H_{c3}^{SU(4)} &=&4 g_t^{-1}.  \nonumber
\end{eqnarray}
We compare the above analytic results with TBA numerical ones in the Fig.
\ref{SU4gt}A, which shows an excellent accuracy for most values of $g_t$.
Notice
that the deviations occur only for less-physical values of $g_t$ due to
small Fermi points. For $g_s=2.0$ and $g_t=1.0$, the discrepancies of $%
H_{c1}^{SU(4)}$ and $H_{c2}^{SU(4)}$ (TBA numerics: 0.3695, 0.9415;
analytic: 0.3697, 0.9386 (200 sites results\cite{SJGu} for comparison: 0.31,
0.93)) are, respectively, 0.05\% and 0.3\%.  $H_{c3}^{SU(4)}$ is
exact. The magnetization of the SU(4) model is shown in Fig.\ref{Mz}A,
which also coincides with the numerical result for 200 sites\cite{SJGu}.

We thank Huan-Qiang Zhou, You-Quan Li and M.T. Batchelor 
for discussions and comments.
ZJY thanks FAPERGS and FAPERJ for partial support. 
AF thanks FAPERGS and CNPq. 
XWG thanks the Australian Research Council for support. 
BC thanks NSF (Natural Science Foundation) of
Zhejiang, China (RC02068) and NSF of China
(10274070). IR thanks PRONEX ans CNPq.

\newpage
\end{multicols}{2}


\begin{references}
\bibitem{Nagaosa}  Y. Tokuro and N. Nagaosa, Science {\bf 288}, 462(2000).

\bibitem{Saitoh}  E. Saitoh {\it et al}., Nature (London) {\bf 410}, 180
(2001).

\bibitem{LiSU4}  Y.Q. Li {\it et al}., Phys. Rev. Lett. {\bf 81},3527
(1998).

\bibitem{LiSU4BA}  Y.Q. Li {\it et al}., Phys. Rev. B {\bf 60}, 12781(1999).

\bibitem{Yamashita}  Y. Yamashita {\it et al}., Phys. Rev. B {\bf 61}, 4012
(2000).

\bibitem{SJGu}  S.J. Gu and Y.Q.Li, Phys. Rev. B {\bf 66}, 092404 (2002).

\bibitem{Mila} F. Mila {\it et al.}, Phys. Rev. Lett. {\bf 82}, 3697
(1999).

\bibitem{Azaria}  P. Azaria {\it et al.}, Phys. Rev. Lett. {\bf 83}, 624
(1999).

\bibitem{Frischmuth} B. Frischmuth {\it et al.}, Phys. Rev. Lett. {\bf 82}, 835
(1999).

\bibitem{Haldane}  F.D.M. Haldane, Phys. Lett. A {\bf 93}, 464 (1983).

\bibitem{Oshikama}  M.Oshikama {\it et al.}, Phys. Rev. Lett.
{\bf 78}, 1984 (1997).

\bibitem{FracMz}  K.Kodama {\it et al.}, Science {\bf 298}, 395 (2002);
K.Onizuka {\it et al}., J. Phys. Soc. Jpn. {\bf 69}, 1016 (2000); G.
Misguich {\it et al}., Phys. Rev. Lett. {\bf 87}, 097203 (2001);
B.S. Shastry and B. Sutherland, Physica B {\bf 108}, 1069 (1981).

\bibitem{Slonczewski}  J.C. Slonczewski, Phys. Rev. {\bf 110}, 1341 (1958).

\bibitem{Sutherland}  B. Sutherland, Phys. Rev. B {\bf 12}, 3795 (1975).

\bibitem{string}  M. Takahashi, Prog. Theor. Phys. {\bf 46}, 401 (1971).

\bibitem{Y-Y}  C.N. Yang and C.P. Yang, J. Math. Phys. {\bf 10}, 1115(1969).

\bibitem{WH}  M.G. Krein Usp. Mat. Nauk {\bf 13}, 3 (1958).


\end{references}
\end{document}